\documentclass[conference]{IEEEtran}
\makeatletter
\def\ps@headings{%
\def\@oddhead{\mbox{}\scriptsize\rightmark \hfil \thepage}%
\def\@evenhead{\scriptsize\thepage \hfil \leftmark\mbox{}}%
\def\@oddfoot{}%
\def\@evenfoot{}}

\makeatother
\usepackage{graphicx}
\usepackage{times}
\usepackage{graphics}
\usepackage{framed,lipsum}
\usepackage{url}
\usepackage{subfig}
\usepackage{epsfig}
\usepackage{caption}
\usepackage{varwidth}
\usepackage{algorithm}
\usepackage{algorithmicx,algpseudocode}

\newcommand{\CASE}[1]{\STATE \textbf{case} #1\textbf{:} \begin{ALC@g}}
\newcommand{\ENDCASE}{\end{ALC@g}}

\newcommand{\DEFAULT}{\STATE \textbf{default:} \begin{ALC@g}}
\newcommand{\ENDDEFAULT}{\end{ALC@g}}
\newcommand{\DEFAULTLINE}[1]{\STATE \textbf{default:} }
\usepackage{color}
\usepackage{multirow}
\usepackage{multicol}
\usepackage{hhline}
\usepackage{amsthm}
\usepackage{comment}
\usepackage{tabu}

\theoremstyle{plain}

\theoremstyle{definition}

\theoremstyle{remark}

\IEEEoverridecommandlockouts
\begin{document}

\title{Toward a Programmable FIB Caching Architecture}
\author{
Garegin Grigoryan \\ grigorg@clarkson.edu  \\Clarkson University
\and Yaoqing Liu \\liu@clarkson.edu \\
Clarkson University
}

\makeatletter
\def\ps@IEEEtitlepagestyle{
	\def\@oddfoot{\mycopyrightnotice}
	\def\@evenfoot{}
}
\def\mycopyrightnotice{
	{\footnotesize
		\begin{minipage}{\textwidth}
			978-1-5090-6501-1/17/\$31.00~\copyright~2017 IEEE
			
		\end{minipage}
	}
}

\maketitle

\pagestyle{empty}

\begin{abstract}

The current Internet routing ecosystem is neither sustainable nor economical. More than 711K IPv4 routes and more than 41K IPv6 routes exist in current global Forwarding Information Base (FIBs) with growth rates increasing. This rapid growth has serious consequences, such as creating the need for costly FIB memory upgrades and increased potential for Internet service outages. And while FIB memories are power-hungry and prohibitively expensive, more than 70\% of the routes in FIBs carry no traffic for long time periods, a wasteful use of these expensive resources. Taking advantage of the emerging concept of programmable data plane, we design a programmable FIB caching architecture to address the existing concerns. Our preliminary evaluation results show that the architecture can significantly mitigate the global routing scalability and poor FIB utilization issues. 

\end{abstract}

\section{Introduction}
\label{sec:intro}

The current Internet routing ecosystem is not green nor sustainable. Internet Service Providers (ISPs) leverage backbone routers with prohibitively expensive linecard memories to forward Internet traffic in line rate. Backbone routers are power-hungry machines. TCAM, the heart of almost every hardware switch/router, is a unique memory that can do parallel lookups on multiple fields in one clock cycle. TCAMs are considerably power hungry and also considered to be the most expensive component on commodity switches and line cards~\cite{katta2014infinite}-100X more power-consuming ($\mathtt{\sim}$15W/Mb) and 400X more expensive($\mathtt{\sim}$\$350/Mb) per Mbit than RAM-based memories. More importantly, a TCAM chip occupies 6 times or more board space than that of a SRAM chip~\cite{meiners2011split}, which leads to TCAM chips being even more expensive. Meanwhile, due to various network operations, such as traffic engineering, multi-homing, prevention of BGP hijacking, etc., the Internet routing table size has been out-pacing the upgrade cycle of router/switch hardware. Today's Internet routing system has more than 711K IPv4 routes, plus more than 41K IPv6 routes that grow within a potentially much larger space  (128 bits) than IPv4 (32 bits). The growth of Internet routing table may result in router memory overflow, unstable network connectivity, and unreliable or even unavailable network service. 
Despite the facts above, current routing practices are not economical and bear a high waste of expensive and power-hungry networking resources. Existing studies~\cite{sarrar2012leveraging} have shown that the Internet traffic demonstrates a pattern with high skewness, namely, a small set of popular routes carry most of the observed Internet traffic. 

The emerging Protocol Independent Switch Architecture (PISA) architecture~\cite{bosshart2014p4} provides reconfigurable switch chips that process packets as fast as the fastest  switches on the market, while P4 language~\cite{p4} offers high-level language support to flexibly configure forwarding tables on the chips, rather than using a fixed-function OpenFlow-like approach. Additionally, they provide many new programmable capabilities such as shared metadata between different stages, on-chip counters/registers, packet recirculation, etc. In our design, we leverage both PISA architecture and P4 language to program and reconfigure data planes to design and implement the system. 

We have made the following contributions in this work: (1) We designed an architecture to divide a full forwarding table into three sub-tables into three types of memories, TACM, SRAM and DRAM. TCAM table contains popular routes, SRAM contains unpopular ones and DRAM contains a full table and flags to track which routes are in TCAM or SRAM; (2) Based on the availability of memory resources on a specific target, we can dynamically allocate different sizes of memory for each type of routes. As a result, the cost of the overall hardware and the amount of power consumption in the designed system can be significantly reduced compared with existing solutions, where TCAM is primarily used to hold the full forwarding table and more than 70\% of the routes are not used at all for a long period. 

\section{Design Overview}
\label{sec:overview}

\begin{figure}
	\begin{center}
		\includegraphics[width=1\columnwidth]{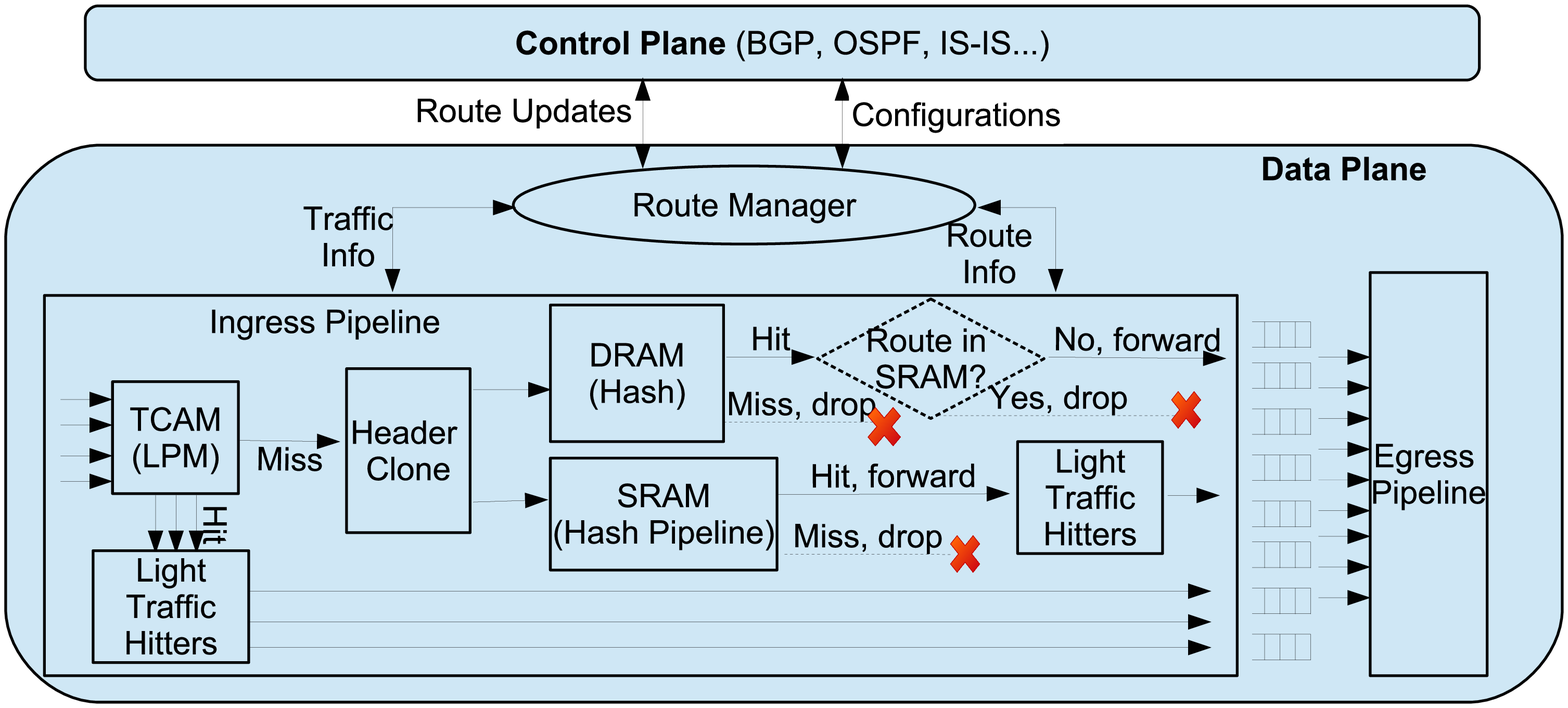}
		\caption{\label{fig:architecture}Architecture}
	\end{center}
	\vspace{-8mm}
\end{figure}

Figure~\ref{fig:architecture} illustrates conceptually the architecture of the proposed Programmable FIB Caching System (PFCS). The control plane is responsible for feeding route updates and issuing configuration commands to the programmable data plane. In the data plane, PFCS extends P4 by introducing an external Route Manager module. The Ingress pipeline consists of three types of memories{\textemdash}TCAM, SRAM and DRAM{\textemdash}to store FIB routes to forward data packets. The TCAM, SRAM and DRAM will store popular routes, unpopular routes and rarely used routes, respectively. Initially, the routes in each memory will be allocated by Route Manager, which can be implemented by a micro-controller that may include a CPU, a DARM memory, etc. Route Manager controls the full FIB in DRAM and is in charge of assigning routes to individual memories based on collected FIB updates and configurations from the control plane as well as traffic/routes statistics from Ingress and Egress Pipelines. Two traffic-driven Light Traffic Hitters modules for TCAM and for SRAM filter out routes that have relatively high traffic loads. The routes in the two modules will be directly used as candidate victim routes when either memory is full, which triggers the need to evict an old route with a new one. 

The overall workflow is as follows. First, all traffic goes through TCAM where Longest Prefix Matching (LPM) is enforced and more than 99.9\% of the traffic (according to our preliminary work) ends up with cache hits. The cache-hit traffic then heads to Light Traffic Hitters module, where a small set of active routes with the lightest traffic load in TCAM are kept; they will be used as victim routes to be replaced by more popular ones. Meanwhile, less than 0.1\% of the traffic is cache missed, which heads to a Header Clone module. This module clones the header of each packet; one copy heads to the SRAM cache, and simultaneously the other heads to the DRAM, where a full forwarding table resides. Another important functionality of DRAM is that it tracks which routes reside in the TCAM cache and which ones reside in the SRAM cache. This feature allows us to detect two possible events. First, if a packet is cache missed again in the SRAM cache, its forwarding process terminates. Since the DRAM keeps track of which memory contains which routes, and thus is aware of the mismatch, it will forward the packet out and simultaneously update the SRAM cache to include a new route against the cache missed data packet. Alternatively, if a packet ends up with a cache hit in the SRAM cache and the DRAM also recognizes it through its route records, the forwarding process of the cloned one terminates in DRAM but continues in the SRAM cache. This design adds minimal redundancy for header duplication but reduces the forwarding latency when a further cache miss occurs in the SRAM cache. 
%
%

\section{Preliminary Results}
\label{sec:evaluation}

\begin{figure}
	\begin{center}
		\includegraphics[width=0.7\columnwidth]{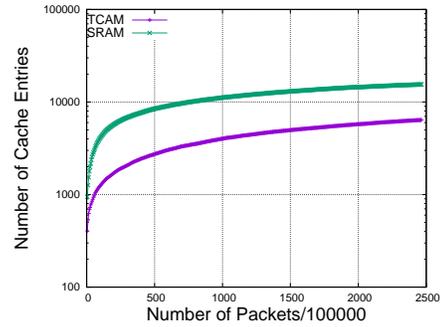}
		\caption{\label{fig:cacheentries}Number of Entries in Caches}
	\end{center}
	\vspace{-8mm}
\end{figure}

\begin{figure}
	\begin{center}
		\includegraphics[width=0.7\columnwidth]{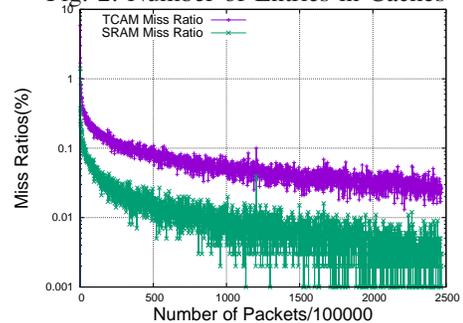}
		\caption{\label{fig:missratio}Miss Ratios}
	\end{center}
	\vspace{-9mm}
\end{figure}

We have partially simulated the architecture using both C and P4 programming languages.  We use a realistic tier-1 Internet trace that contains 245 million packets to carry out the experiments and a real Internet routing table that has 560K entries to populate the needed cache entries.  Initially, we set the cache sizes 10K and 20K for TCAM and SRAM, respectively. We first measured the number of non-overlapping routes for both TCAM and SRAM caches as shown in Figure~\ref{fig:cacheentries}, and then obtained their cache miss ratios as shown in Figure~\ref{fig:missratio}. Figure~\ref{fig:cacheentries} shows that only 6K and 16K routes among 560K full routing entries are needed to carry most of the traffic. Other more than 500K routes can be stored in a relatively slow and cheap DRAM memory. Figure~\ref{fig:cacheentries} demonstrates the caching miss ratios of every 100K packets, we can observe that most of the time less than 0.1\% of the traffic is not captured by TCAM cache and less than 0.01\% is cache missed in SRAM and thus is forwarded by DRAM. In other words, if the total traffic rate is 1 Tbps at the router, then the DRAM only needs to handle less than 100 Mbps of the traffic, which is an easy task for a normal DRAM. Compared with an existing router without caching capabilities, both hardware costs and energy consumption will be significantly reduced.   

\section{Conclusion and Future Work}
\label{sec:conclusion}

We leverage the emerging new paradigm of programmable data plane to design a scalable FIB caching architecture, which has shown promising performance based on some preliminary results using realistic traffic traces and routing tables. Our future work includes designing efficient data structures and algorithms for each module to quickly detect light traffic hitters and to effectively handle FIB updates derived from the control plane. We also plan to conduct the experiments in a real hardware-based programmable switch and evaluate the exact savings for both hardware costs and energy consumption. 
\bibliographystyle{IEEEtran}
\bibliography{net}
\end{document}